\documentclass{aa}
\usepackage{graphicx}

\def\jh{\mbox{$\rm (J-H)$}}

\def\mMJ{\mbox{$\rm (m-M)_J$}}
\def\ebv{\mbox{$\rm E(B-V)$}}
\def\ejh{\mbox{$\rm E(J-H)$}}

\def\rc{\mbox{$\rm R_{core}$}}

\def\ms{\mbox{$\rm M_\odot$}}
\def\ds{\mbox{$\rm d_\odot$}}

\def\jj{\mbox{$\rm J$}}

\def\ks{\mbox{$\rm K_S$}}

\def\mobs{\mbox{$\rm m_{obs}$}}
\def\mtot{\mbox{$\rm m_{tot}$}}

\def\wA{\mbox{$\rm W\,306.5-0.6$}}
\def\wB{\mbox{$\rm W\,307.0-0.7$}}

\begin{document}

\title{A low-absorption disk zone at low Galactic latitude in Centaurus}

\author{E. Bica\inst{1} \and Ch. Bonatto\inst{1} \and Bas\'\i lio X. Santiago\inst{1}
\and Leandro O. Kerber\inst{1} }

\offprints{Ch. Bonatto - charles@if.ufrgs.br}

\institute{Universidade Federal do Rio Grande do Sul, Instituto de F\'\i sica, CP\,15051, 
Porto Alegre 91501-970, RS, Brazil\\
\mail{} }

\date{Received --; accepted --}

\abstract{We investigate the properties of two stellar concentrations in a low-absorption 
disk zone in Centaurus, located respectively at $\ell=306.47^{\circ}$, $b=-0.61 ^{\circ}$, 
and $\ell=307.01^{\circ}$, $b=-0.74 ^{\circ}$. The present analysis is based mostly on 
2MASS photometry, as well as optical photometry. Based on colour-magnitude diagrams and 
stellar radial density profiles, we show that these concentrations are not open star 
clusters. Instead, they appear to be field stars seen through a differentially-reddened 
window. We estimate that the bulk of the stars in both stellar concentrations is located at 
$\sim1.5$\,kpc from the Sun, a distance consistent with that of the Sgr-Car arm in that 
direction. This low-absorption window allows one to probe into distant parts of the disk 
besides the Sgr-Car arm, probably the tangent part of the Sct-Cru arm, and/or the far side 
of the Sgr-Car arm in that direction. The main sequence associated to the Sgr-Car arm is 
reddened by $\ebv\sim0.5$, so that this window through the disk is comparable in reddening to 
Baade's window to the bulge. We also investigate the nature of the open cluster candidate 
Ru\,166. The presently available data do not allow us to conclude whether Ru\,166 is an 
actual open cluster or field stars seen through a small-scale low-absorption window.

}

\titlerunning{A low-absorption zone in Centaurus}
\authorrunning{E. Bica et al.}

\maketitle

\section{Introduction}
\label{intro}

Low-absorption zones are fundamental for optically probing distant parts of the Galaxy in 
order to understand its structure, abundance and kinematics. Baade's (\cite{Baade1946}) 
window has played a major role in the studies of the Galactic bulge (e.g. Terndrup 
\cite{Tern1988} and McWilliam \& Rich \cite{MWR1994}). In addition to Baade's window two 
other low-absorption bulge windows, Sgr\,I and Sgr\,II, have been described in Baade 
(\cite{Baade1963}) and Lloyd Evans (\cite{LLE1976}). More recently, Dutra, Santiago \&
Bica (\cite{DSB2002}) identified a more centrally located window in the bulge whose position
makes it a potential source of information complementary to Baade's window. 

Along the Milky Way, it would be valuable to isolate low-absorption windows 
at low galactic latitudes to sample more distant parts of the disk, e.g. for optical 
abundance studies of hot stars (Daflon et al. \cite{DCS2001}, \cite{DCS2003}, and Mathys et 
al. \cite{Mathys2002}).

The present target direction ($\ell\approx307^\circ$) has been studied in Galactic
structure surveys (Georgelin \& Georgelin \cite{GG1976}; Russeil \cite{Rus2003}).
This direction crosses the Sgr-Car arm at 1.6\,kpc from the Sun, and tangentially
intercepts the Sct-Cru arm at 6.6\,kpc, and finally crosses the
far side of the Sgr-Car arm at 12\,kpc. To the south of the studied region, there
occurs the extended diffuse \ion{H}{ii} region Cederblad\,122, also know as m Centauri 
nebula (Georgelin \& Georgelin \cite{GG1970}). Note that the star m Centauri (HD\,116243) 
is a close-by G\,6\,II star, at a distance to the Sun of 79\,pc according to SIMBAD,
thus unrelated to the nebula. The presence of an emission nebula, revealing star 
formation, projected close to the direction we are studying is important, since
our target region might be an overall star-forming region in the Sgr-Car spiral 
arm (or an association). The emission nebula 
has a kinematic distance of 1.75\,kpc, thus compatible with the Sgr-Car arm. The 
central coordinates of this nebula are $\ell=306.64^\circ$ and $b=-1.39^\circ$ and 
J2000.0 $\alpha=13^h25^m24^s$ and $\delta=-64^\circ\,01\arcmin\,00\arcsec$.

We analyze in the present study a high stellar surface density region encompassing
$1.9^\circ\times0.9^\circ$, which detaches in Sky Survey plates and can be seen
in wide-angle Milky Way photographs in Centaurus. The central coordinates of this region 
are J2000.0 $\alpha=13^h20^m17^s$ and $\delta=-63^\circ\,36\arcmin\,46\arcsec$, which 
correspond to $\ell=306.13^\circ$ and $b=-0.92^\circ$.  Low absorption in the region
$\ell\approx305^\circ$\ had already been reported (FitzGerald \cite{Fitz68}), where it
was found that the excess \ebv\ is largely $\sim0.5$\ or less to $\rm r\sim4$\,kpc. 
The main goal of the present paper 
is to unveil the nature of this region, whether it is a star cluster or association resulting from an 
enhanced star-formation event and/or a low-absorption zone (window). This kind of analysis 
has become possible with the advent of the {\em Two Micron All Sky Survey} (2MASS, Skrutskie 
et al. \cite{2mass}) photometry, whose uniformity and spatial coverage allow one to probe 
large areas in the sky using the near-infrared. 

In the target area there are two conspicuous stellar concentrations located
at J2000.0 $\alpha=13^h 23^m 01^s$ $\delta=-63^{\circ} 16\arcmin 03\arcsec$ ($\ell = 
306.47^{\circ}$, $b = -0.61 ^{\circ}$) and $\alpha=13^h 27^m 56^s$ $\delta=-63^{\circ} 
19\arcmin 49\arcsec$ ($\ell = 307.01^{\circ}$, $b = -0.74 ^{\circ}$). Optically
they resemble rich star clusters, in particular due to the contrast in stellar surface 
density between the central concentrations and surroundings (Fig.~\ref{fig1}). 
We study in detail these two regions by means of colour-magnitude diagrams (CMDs) and 
radial density of stars. Considering the positions and optical appearance, they might 
be identified as NGC\,5120 and NGC\,5155, originally described as rich star clusters 
in the New General Catalogue (Dreyer 1888). Note that the original coordinates 
(precessed to J2000.0) are shifted by $21\arcmin$ and $12\arcmin$ towards SE, respectively 
for NGC\,5120 and NGC\,5155; however, their angular diameters of $\approx20\arcmin$ basically 
encompass the shifts. We also point out that NGC\,5120 has been identified as the open 
cluster candidate Ruprecht\,166 (hereafter Ru\,166) or as ESO\,96-SC\,11 (Lauberts \cite{Lau1982}). 
In the latter study, NGC\,5155 has been identified as the small concentration of stars 
ESO\,96-SC\,13. 

An alternative designation for features resembling field windows was introduced by Dutra 
et al. (2002) which includes the galactic longitude and latitude in the name. 
Accordingly, hereafter the first feature described above will be designated as 
\wA\ and the second as \wB. For practical reasons we refer to them
as windows, in the sense of a case similar to that of Baade's window. Indeed,
\wA\ and \wB\ are surrounded by dark nebulae such as H\,304.7-0.3
to the West, H\,306.2-0.3, Fest-1\,239, H\,306.9-0.1 to the North, Fest-1\,245
and H\,307.2-1.0 to the East, and Fest-2\,168 to the South (Dutra \& Bica 
\cite{DB2002} and references therein).

We employ in the present study near-infrared photometry from 2MASS as well as optical 
data from the Guide Star Catalogue to explore the nature of \wA\ and \wB. In 
Section~\ref{CMDS} we present the data extraction from optical and infrared databases,
discuss the near-infrared CMDs, the radial density profiles and estimate parameters for 
the features \wA\ and \wB. In Sect.~3 we investigate the nature of the open cluster 
candidate Ru\,166, which is located midway and slightly south of the features \wA\ and 
\wB\ (Fig.~\ref{fig1}). Concluding remarks are given in Sect.~\ref{conclu}.

\section{The nature of the W\,306.5-0.6 and W\,307.0-0.7 features}
\label{CMDS}

We investigate the nature of \wA\ and \wB\ by means of optical and near-infrared images, 
radial density  profiles and CMDs.

\subsection{Optical data}
\label{opticaldata}

In the optical we  are interested in the detection of as many stars as possible within the 
sampled regions. We employed the Guide Star Catalogue (GSC\footnote{\em 
http://www-gsss.stsci.edu/gsc/GSChome.htm}, version 2.2) to extract the positions of all 
stars contained within an area with $40\arcmin$ in radius centered midway between \wA\ and 
\wB. As a result, we obtained the positions and magnitudes of 334\,424 stars, including all 
the available bands in the Guide Star Catalogue.

\begin{figure} 
\resizebox{\hsize}{!}{\includegraphics{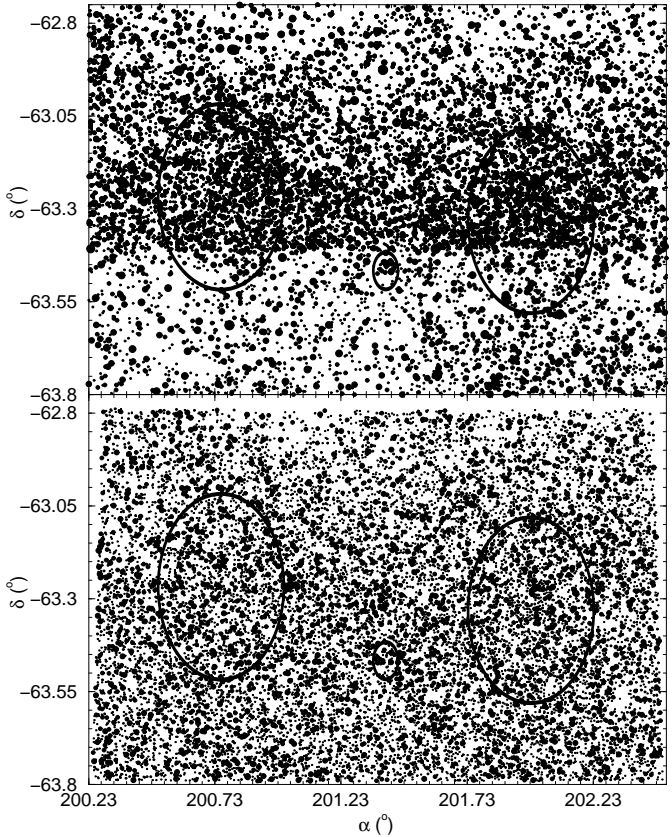}}
\caption[]{Top panel: GSC map of the target region showing the positions of the stars 
with magnitude $V\leq 15$. The features \wA\ (right stellar concentration) and \wB\ (left) 
are marked off. The region occupied by the open cluster candidate Ru\,166 
($\alpha\approx201.4^\circ, \delta\approx-64.5^\circ$) is marked off as well. Note an
apparent non-uniformity in the GSC sampling for $\delta\leq-63.4^\circ$. Bottom panel: same 
as above for stars with 2MASS magnitude $\jj\leq 13$; note that the number density of
stars decreases for $\delta\geq-63.05^\circ$. The area covered in both panels is 
$2.2^\circ\times1^\circ$.}
\label{fig1}
\end{figure}

We show in Fig.~\ref{fig1} (top panel) a GSC\,II map with dimensions $2.2^\circ\times1^\circ$ 
of the region containing the features \wA\ and \wB. At first glance, these structures resemble 
actual star clusters, in particular by the contrast in the stellar surface density between both 
features and surrounding regions. Note that there appears to be a non-uniformity in the GSC 
sampling of stars, since the lower third of the GSC image ($\delta\leq-63.4^\circ$) is 
systematically less dense. 

\begin{figure} 
\resizebox{\hsize}{!}{\includegraphics{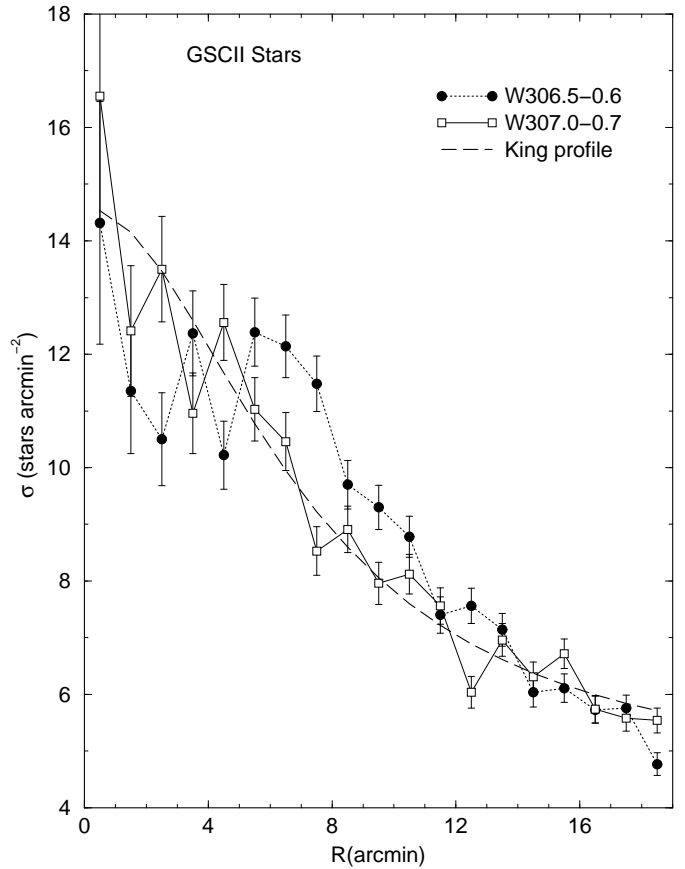}}
\caption[]{Radial distribution of stars around the center of \wA\ (filled circles) and \wB\ 
(open squares) obtained from the optical photometry (GSC). The radial distribution of \wB\ 
can be fitted by a King profile (dashed line). Bars correspond to Poisson fluctuations.}
\label{fig2}
\end{figure}

To investigate how the stars are distributed throughout the features \wA\ and \wB\, we construct 
the optical stellar distribution profiles, i.e., the radial density of stars per square arcmin, 
measured in radial annuli around the center of each feature. The resulting radial density profiles 
are shown in Fig.~\ref{fig2}. Surprisingly, the profile of \wB\ can be reasonably fitted by a 
King (\cite{King1962}) profile, $\sigma(r)=\frac{\sigma_0}{1+(r/\rc)^2}$, possibly mimicking an 
actual star cluster (Sect.~\ref{nirphot}). However, the same does not apply to \wA.

\subsection{Near-infrared photometry}
\label{nirphot}

Near-infrared J and H photometry have been extracted from the 2MASS All-Sky Data 
Release\footnote{\em http://www.ipac.caltech.edu/2mass/releases/allsky/}. The 2MASS 
photometry has proven to be an excellent tool to be applied in  analyses of the physical 
structure and luminosity/mass functions of star clusters by means of the radial 
distribution of stars and CMDs (see, e.g. Bonatto, Bica, \& Santos Jr. \cite{BBJ2004},
Bonatto, \& Bica \cite{BB2003} and Bica, Bonatto, \& Dutra \cite{BiBoDu2004}). Star
clusters and galactic structure have also been addressed with 2MASS data (e.g. Bica, 
Bonatto, \& Dutra \cite{BiBoDu2003}).

In the bottom panel of Fig.~\ref{fig1} we plot the positions of stars with magnitude 
$\jj\leq 13$ taken fom 2MASS, which are located in the target area. Contrary to what is 
observed in the optical image (top panel of Fig.~\ref{fig1}), in the near-infrared the 
features \wA\ and \wB\ do not present a contrast in terms of stellar surface density with 
respect to the surroundings. However, a slight decrease in the surface density can be
seen for declinations $\delta\geq-63.05^\circ$.

For both \wA\ and \wB\ we made circular extractions of the stars contained within an area of 
20\arcmin\ in radius, centered on the corresponding coordinates given in Sect.~\ref{intro}. 
Stars in comparison fields have been extracted in areas of the same radius as that used for 
the features, to the North and South of both features, with center to center distances of 
60\arcmin. Extractions have been performed using the 2MASS tool at the Infrared Science Archive 
(IRSA\footnote{\em http://irsa.ipac.caltech.edu/cgi-bin/Gator/nph-dd}). 

\begin{figure} 
\resizebox{\hsize}{!}{\includegraphics{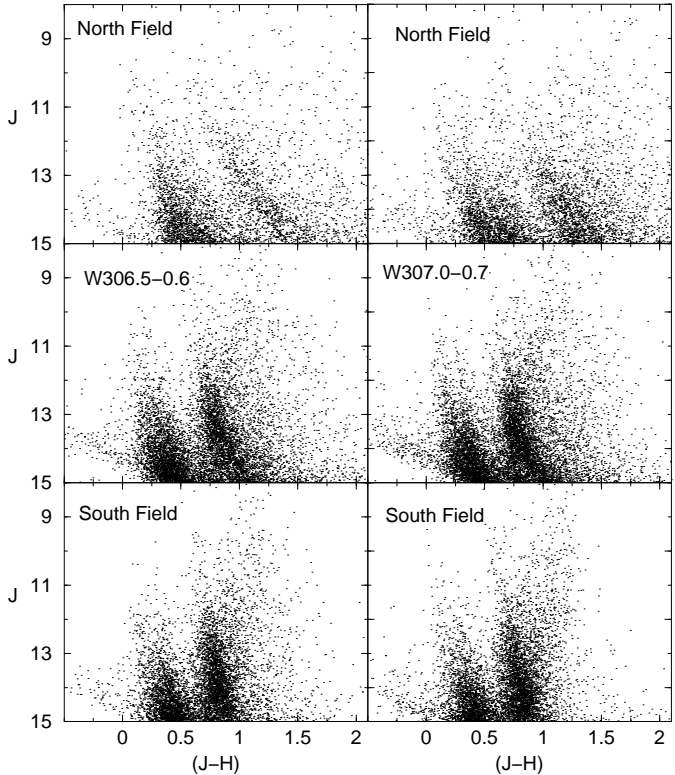}}
\caption[]{Left panels: $\jj\times\jh$ CMD of \wA\ and offset fields extracted
from 2MASS within an area of 20\arcmin\ in radius. North and South offset fields 
are located at 60\arcmin\ from the central coordinates. Right panels: same as the left 
panels for \wB.}
\label{fig3}
\end{figure}

The resulting 2MASS $\jj\times\jh$ CMDs are shown in Fig.~\ref{fig3} for the \wA\ (left 
panels) and \wB\ (right panels). The middle panels deal with the feature extraction, 
while the CMDs of the North and South offset fields are shown in the top and bottom panels, 
respectively. The North field of both features is clearly more reddened than the features 
themselves, while the South fields are similarly reddened than the features. This fact can 
be accounted for by the North fields being projected closer to the galactic plane than the South 
fields. We also checked CMDs involving \ks\ and found that those involving J and H were 
better defined. Two nearly vertical extended star sequences (a blue and a red one) appear in 
the feature CMDs. This suggests the presence of relatively young stars, and not an older disk 
stellar population. Both star sequences may result from the upper main sequences of a 
dominant disk stellar population. The difference in colour between the two sequences certainly 
reflects different extinction related to spiral arms at different distances. The blue 
sequence is located at $0.25\leq\jh\leq 0.50$ and the red one is at $0.75\leq\jh\leq 1.00$ 
for both features. There is no compelling evidence for these star sequences to arise from 
star clusters, since the CMDs of the feature regions and comparison fields are similarly 
populated. The resulting feature CMDs may originate from background stars seen through a 
low-absorption zone. However, the presence of an association or an enhanced star-forming 
region related to the spiral arm cannot be ruled out at this point. The blue and red
sequences in the CMDs of \wA\ and \wB\ are wide, indicating differential reddening. In the
north fields the sequences are even wider, suggesting more differential reddening.

\begin{figure} 
\resizebox{\hsize}{!}{\includegraphics{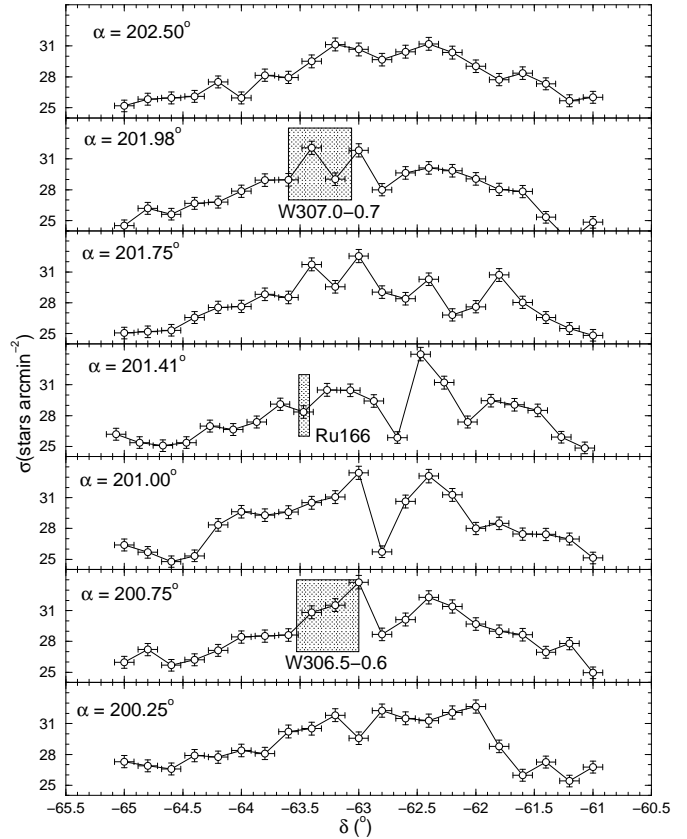}}
\caption[]{Stellar surface density ({$\rm stars\,arcmin^{-2}$}) distribution across the target 
region for stars with 2MASS photometry. The areas corresponding to \wA\ and \wB\ are marked off
by shaded rectangles. The distribution suggests an excess of stars in the areas of \wA\ and \wB\ 
with respect to the surroundings. The location of the small stellar concentration Ru\,166 is as 
well indicated.}
\label{fig4}
\end{figure}

Another evidence in favour of both \wA\ and \wB\ being in a low-absorption 
window can be seen in Fig.~\ref{fig4}, in which we show the stellar surface density
distribution (in terms of {$\rm stars\,arcmin^{-2}$}) across the target area. The surface density 
distribution has been built by calculating the number-density of stars in circular areas of 
5\arcmin\ in radius. The circles are regularly spaced in declination with a center to center 
separation of 12\arcmin. The right ascensions cover nearly all of the target region and, in 
particular, encompass the areas of both \wA\ and \wB\ (as well as the open cluster candidate 
Ru\,166). Both features are located in regions where the density of stars is higher than
the surroundings. Indeed, the stellar surface density distributions fall off for declination 
$\delta\leq-63.5^\circ$ and $\delta\geq-62^\circ$, which is consistent with the presence of
a large-scale window in the area. This is the region which detaches on Sky Survey plates
(Sect.~\ref{intro}) and can be designated as {$\rm W\,306.1-0.9$}.

\begin{figure} 
\resizebox{\hsize}{!}{\includegraphics{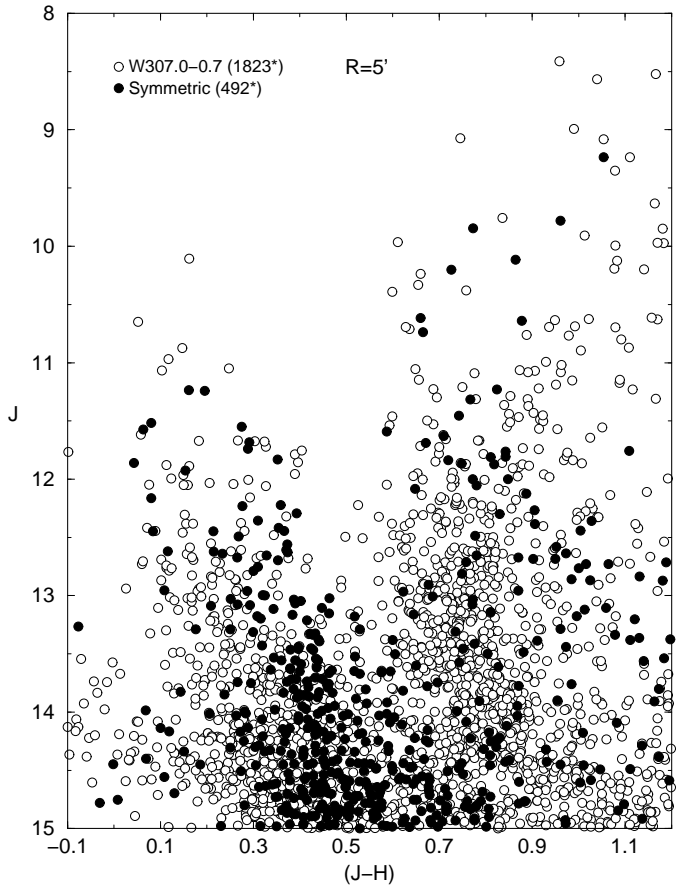}}
\caption[]{CMD of fields symmetrically located with respect to the Galactic plane.
Filled circles: stars extracted from a field at $\ell=307.01^\circ$ and $b=+0.74^\circ$.
Empty circles: stars from the \wB\ feature. For clarity only stars in the central 
5\arcmin\ of each field are considered.}
\label{fig5}
\end{figure}

To further test the reddening gradient scenario we extracted photometry for stars in a 
symmetric field, to the north of the Galactic plane ($b=+0.74^\circ$), with respect to 
the feature \wB. The coordinates of this symmetric field are J2000.0 $\alpha=13^h26^m07^s$ 
and $\delta=-61^\circ\,51\arcmin\,25\arcsec$, which correspond to $\ell=307.01^\circ$ and 
$b=+0.74^\circ$. The resulting CMD is shown in Fig.~\ref{fig5}, superimposed on the CMD 
of \wB. For clarity purposes we restricted the CMDs to the central 5\arcmin\ in both 
fields. There is a severe depletion of stars in the symmetric field (492 stars) with
respect to \wB\ (1823 stars), in the magnitude range $8\leq\jj\leq15$. The blue sequence 
is present in the symmetric field, but more reddened than the corresponding one in 
\wB. Finally, the number of stars in the red sequence of the symmetric field is much 
reduced with respect to  that in \wB. The average reddening parameters for the symmetric 
field are given in Table~\ref{tab1} (see Sect.~\ref{fiduc} for details on reddening 
calculations). With respect to \wB, the blue star sequence in the CMD of the symmetric
field is $\Delta A_V\sim1$\,mag more reddened, again consistent with the large-scale 
window (W\,306.1-0.9) scenario (Sect.~\ref{intro}). 

\subsection{Near-infrared radial profiles}
\label{nirp}

\begin{figure} 
\resizebox{\hsize}{!}{\includegraphics{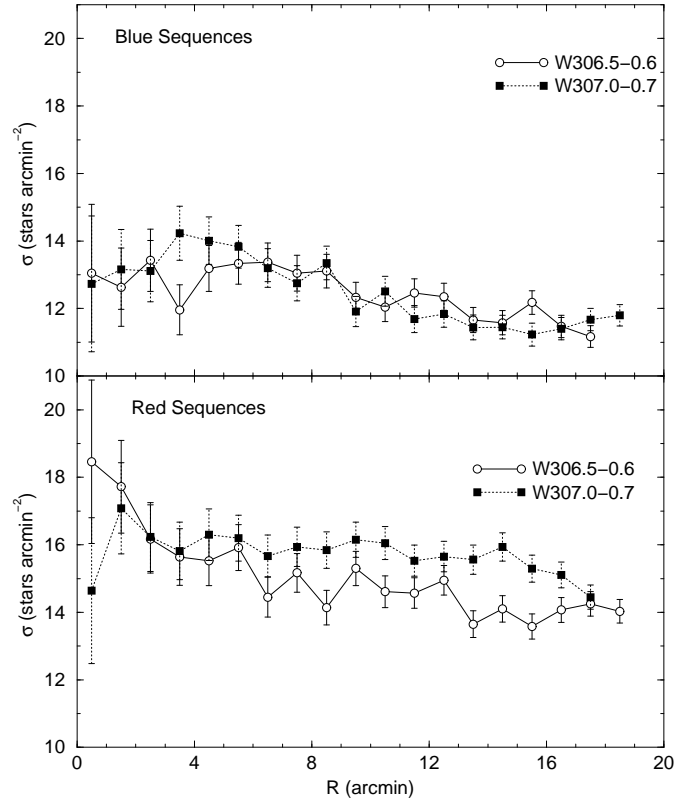}}
\caption[]{Radial distribution of stars from the 2MASS point source catalogue. 
Top panel: stars extracted from the blue CMD sequences. Bottom panel: stars 
from the red sequences. Bars are Poisson uncertainties.}
\label{fig6}
\end{figure}

We show in Fig.~\ref{fig6} the near-infrared radial distribution of stars for \wA\ and \wB, 
which include all stars in the J, H or \ks\ extractions. The distributions correspond to 
the radial number density of stars measured in annuli of radius $\Delta r=1\arcmin$. The 
top panel in Fig.~\ref{fig6} shows the radial distributions for the stars in the blue sequences 
 ($\jh\approx0.1 - 0.5$) seen in the feature CMDs (Fig.~\ref{fig3}), while in the  bottom panel 
are those of the red sequences  ($\jh\approx0.6 - 1.0$).  Within errors, the density profiles are similar in both blue and red CMD sequences 
and much shallower than those in the optical (Fig.~\ref{fig2}), not resembling the profile 
expected from an actual star cluster (see, e.g. M\,67, Bonatto \& Bica \cite{BB2003}, and 
NGC\,188, Bonatto, Bica, \& Santos Jr. \cite{BBJ2004}). Since we are dealing with near-infrared 
data, dust attenuation is less pronounced than in the optical and, consequently, the number of
faint stars detected increases. Recall that peak densities of the optical profiles in
Fig.~\ref{fig2} are slightly different from those in the near-infrared (Fig.~\ref{fig6}).
This arises essentially from the decomposition in two colour sequences in the near-infrared (probably 
different spiral arms), together with slightly different faint magnitude cutoffs. Stars outside
the near-IR colour filters are discarded. Thus, $\jj\sim15$\ and $\rm V\sim15$\ correspond 
in the optical and near-IR to nearly the same stars at the limit, A\,0-F\,8 for a distance of
$\sim1500$\,pc and $\ebv\sim0.5$.  As a consequence of the increased background, the radial density
profiles should have a lower contrast. Thus, the radial density profile similarity between the 
blue and red sequences in both \wA\ and \wB\ features does not support a cluster nature for any 
of the two stellar concentrations. We conclude that all sequences correspond to differentially 
reddened field stars.

\subsection{Dust emission maps}
\label{dust}

\begin{figure} 
\resizebox{\hsize}{!}{\includegraphics{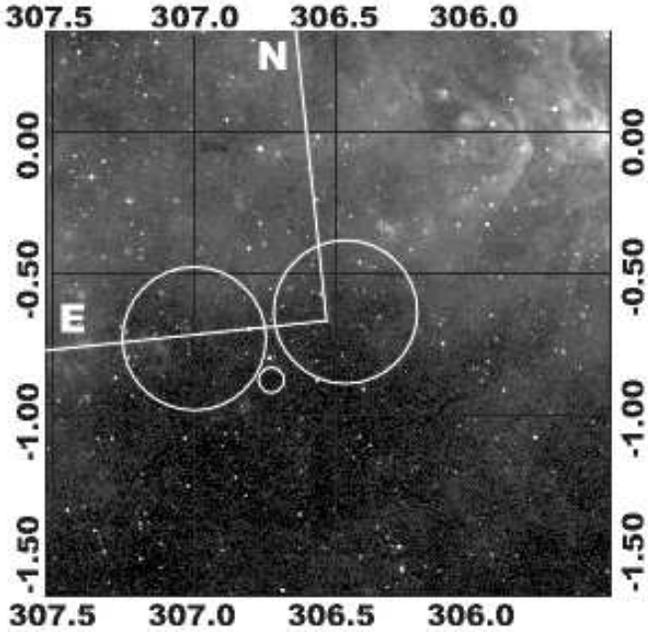}}
\caption[]{$2^\circ\times2^\circ$ MSX image (band\,A, $8.28\,\mu\,m$) of the region 
containing  \wB\ (left circle) and \wA\ (right circle). Galactic north is up and 
Galactic east is to the left. The locus of the open cluster candidate Ru\,166 
($\alpha\approx201.4^\circ, \delta\approx-64.5^\circ$) is marked off as well. The 
equatorial orientation is indicated.}
\label{fig7}
\end{figure}

We show in Fig.~\ref{fig7} an $8.28\,\mu\,m$ (band A) map of the region containing the
features \wA\ and \wB\ obtained from the Midcourse Space Experiment 
(MSX\footnote{\em http://irsa.ipac.caltech.edu/applications/MSX/MSX/ }). In this 
$2^\circ\times2^\circ$ image, a dust emission gradient, increasing to  the north, 
is clearly seen. Notice that this gradient is also present in the near-infrared image 
(Fig.~\ref{fig1}, bottom-panel). Probably this corresponds to a dust layer close
to the Galactic plane  at large distances. Inspection of adjacent fields indicate 
the presence of a low-emission cavity in which both \wA\ and \wB\ are projected. In the 
figure their locus is close to the high-contrast interface. This differential dust 
distribution is probably responsible for the behaviour of the radial stellar distribution 
of \wB\ which, in the optical, mimics a King profile expected in a star cluster 
(Fig.~\ref{fig2}).  This particular dust distribution, i.e. thin in the center and increasing
in density with radius, requires further attention both observationally and theoretically. 
The MSX image suggests the presence of a low dust-emission cavity 
in the area, thus providing low-reddening for optical studies of stars in this direction. 
The cavity has a diameter of $\approx1^\circ$, and can allow low-reddening observations 
to $\approx0.5^\circ$ closer to the galactic plane than in neighbouring directions at 
comparable galactic latitudes.

We also inspected Infrared Astronomical Satellite (IRAS\footnote{\em 
http://irsa.ipac.caltech.edu/applications/IRAS/ISSA/}) images encompassing the target area. 
Despite a lower spatial resolution with respect to MSX, the dust emission gradient 
described above is also present, particularly in the $12\,\mu\,m$ and $100\,\mu\,m$ 
bands.

\subsection{Reddening and distance}
\label{mf}

The discussions presented in the previous sections support the presence of a low-absorption 
disk zone in Centaurus, in which the  apparent stellar concentrations \wA\ and \wB\ stand 
out against the more reddened surroundings. At this point, it may be interesting to estimate 
parameters for the stars in both features, the distance to the Sun in particular. We caution 
that we are sampling a disk direction crossing spiral arms, and thus a mix of disk populations 
is expected, with varying age, metallicity and distance. Nevertheless, the nearly vertical, 
extended CMD blue sequences (Fig.~\ref{fig3}) suggest a dominance in age by stars in a spiral 
arm and its leftover population behind,  i.e. the stars formed during the passage of the 
arm through the disk (Schweizer \cite{Schw76}, see also Vera-Villamizar et al. \cite{Vera2001})
and references therein. Thus, for the sake of estimating parameters, we assume in what follows 
an age of $\sim100$\,Myr for the stars in the blue sequences.

Under the assumption of a relatively young mixture of stellar populations in the features,
we can tentatively estimate parameters, such as distance and reddening, since the line of 
sight interception of a spiral arm displays a distance contrast $\Delta d/d$ which is 
not expected to be large. We use as reference the 100\,Myr solar metallicity 
Padova isochrone from Girardi et al. (\cite{Girardi2002}) computed with the 2MASS J, H and 
\ks\ filters\footnote{Available at {\em http://pleiadi.pd.astro.it}}. The 2MASS 
transmission filters produced isochrones very similar to the Johnson ones, with 
differences of at most 0.01 in \jh\ (Bonatto, Bica \& Girardi \cite{BBG2004}). The 
solar metallicity isochrones have been selected to be consistent with the typical disk 
metallicity. For reddening and absorption transformations we use {$\rm R_V=3.2$}, and 
the relations {$\rm A_J=0.276\times A_V$} and $\ejh=0.33\times\ebv$, according to Dutra, 
Santiago \& Bica (\cite{DSB2002}), and references therein. 

\begin{figure} 
\resizebox{\hsize}{!}{\includegraphics{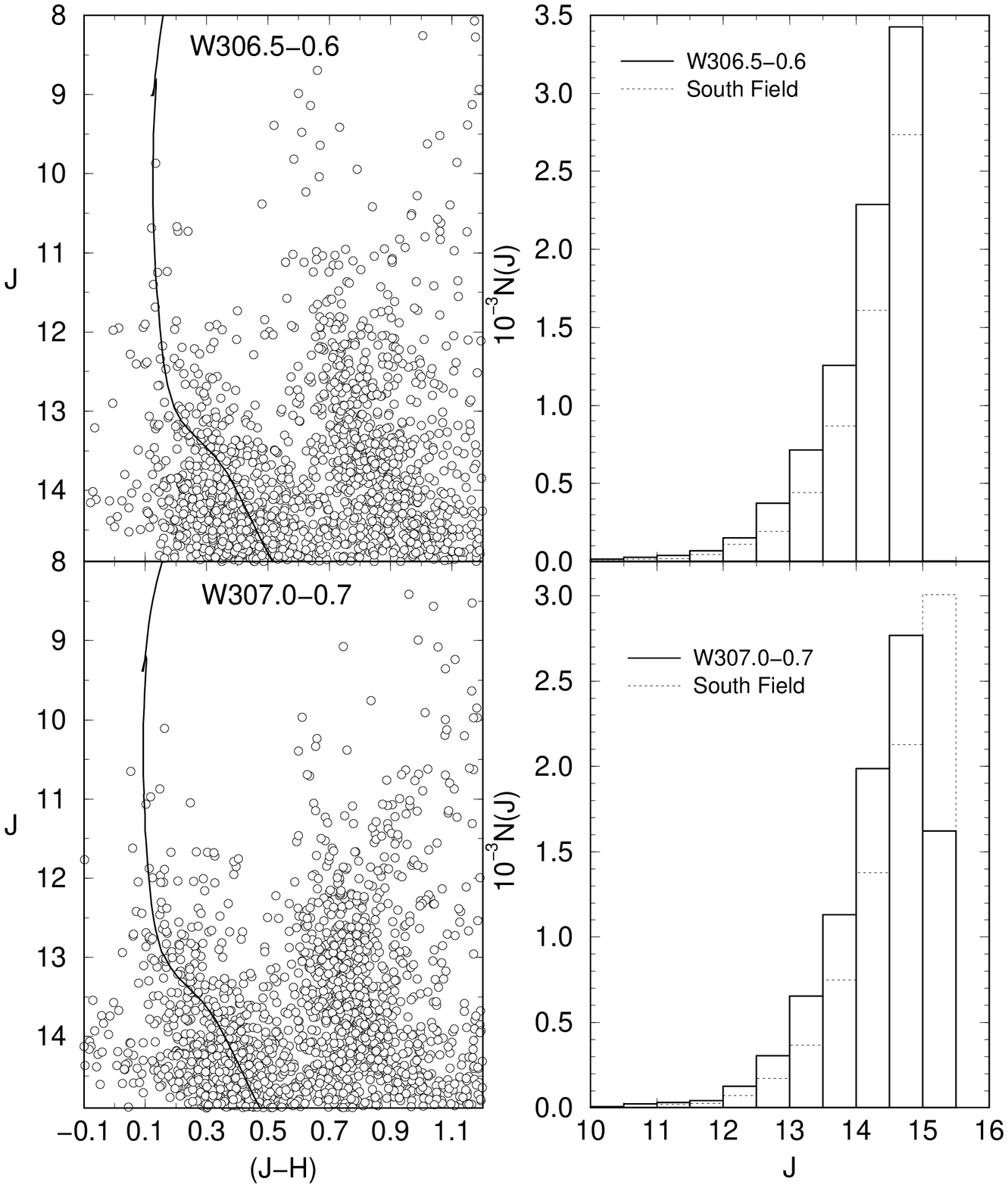}}
\caption[]{Tentative isochrone fits to the CMDs of the features \wA\ (top-left panel) 
and \wB\ (bottom-left panel), with the 100\,Myr, solar metallicity Padova isochrone. 
Parameters for \wA\ are $\ejh=0.18\pm0.02$, $\mMJ=11.2\pm0.1$, and distance to the Sun 
$\ds=1.4\pm0.1$\,kpc; for \wB\ we derive: $\ejh=0.14\pm0.02$, $\mMJ=11.2\pm0.1$, and 
$\ds=1.5\pm0.1$\,kpc. The right panels show histograms for the number of stars per apparent 
\jj\ magnitude bin, for the stars in the blue sequences.}
\label{fig8}
\end{figure}

The resulting tentative fits are shown in Fig.~\ref{fig8} for \wA\ and \wB. For \wA, 
the {\em best-fit} has been obtained with $\ejh=0.18\pm0.02$, $\mMJ=11.2\pm0.1$, and a 
distance to the Sun $\ds=1.4\pm0.1$\,kpc. The {\em best-fit} for \wB\ corresponds to 
$\ejh=0.14\pm0.02$, $\mMJ=11.2\pm0.1$, and  $\ds=1.5\pm0.1$\,kpc. The distance derived for 
both features is consistent with the distance to the Sgr-Car arm in this direction 
(Sect.~\ref{intro}). We provide in the right panels of Fig.~\ref{fig8} histograms for the 
number of stars per apparent \jj\ magnitude bin, for the stars in the blue sequences. For 
both \wA\ and \wB\ features there are excesses in the blue sequences for faint stars, with 
respect to the comparison fields (Fig.~\ref{fig8}, right panels). Finally, since a spatial 
confinement is expected to occur for the bulk of the spiral-arm stars -- which is reflected 
in the CMD morphology of the blue sequences -- we can tentatively derive a mass function 
$\left(\phi(m)\propto m^{-(1+\chi)}\right)$ for the stars in both features. We use the stars 
in the South fields as representative of the stellar-background contribution to the features, 
since these stars are similarly reddened as the features themselves (Fig.~\ref{fig3}). For 
the mass-luminosity relation we use the 100\,Myr Padova isochrone and the distances to the
Sun derived above. We found $\chi=3.3\pm0.2$ for \wA\ and $\chi=3.6\pm0.2$ for \wB. These 
slopes are much steeper than a typical Salpeter mass function ($\chi=1.35$), supporting the 
idea that the features \wA\ and \wB\ are not open clusters, but field stars. Alternatively
we also determined the mass-function slopes for the features without subtracting a
background field. The resulting values turned out to be slightly larger than those derived
above. This result further supports the scenario of \wA\ and \wB\ being composed of field
stellar populations instead of star clusters.

\subsection{Fiducial lines in the CMDs}
\label{fiduc}

\begin{figure} 
\resizebox{\hsize}{!}{\includegraphics{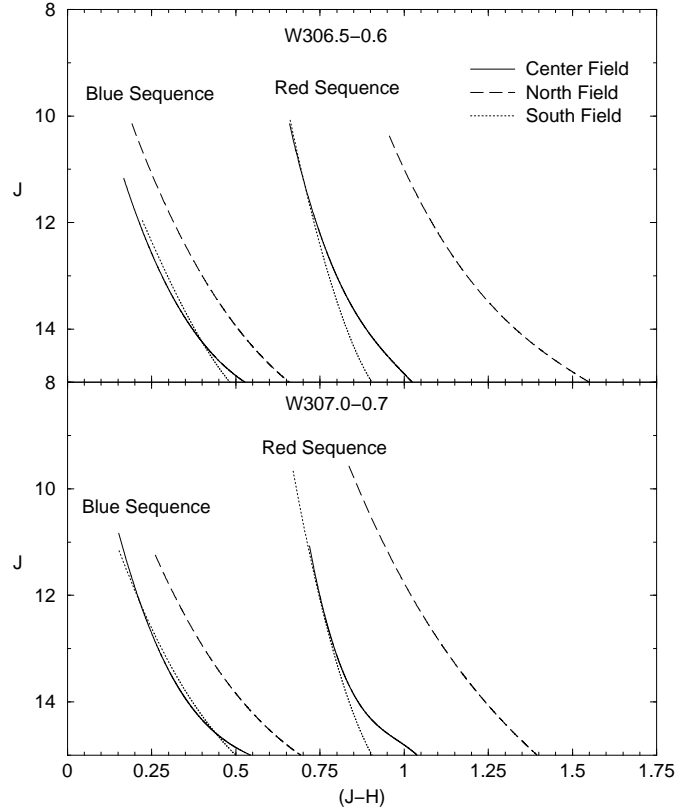}}
\caption[]{Fiducial lines derived from the star sequences observed in the 
2MASS CMDs of \wA\ (top panel) and \wB\ (bottom panel).}
\label{fig9}
\end{figure}

Since we are dealing with field stars, more than one wide MS sequence in colour are
expected to occur on CMDs, owing to differential reddening and occurrence of several
spiral features in the line of sight. In order to derive the main properties of stars, 
it is advisable to first determine the average sequences (fiducial lines) for the observed 
stellar distributions on the CMDs. The fiducial lines for \wA\ and \wB\
and respective north and south fields, are shown in Fig.~\ref{fig9},  where we plot 
the average lines for both blue and red CMD sequences. The fiducial lines have been 
obtained by successively averaging the colours of sets of 100 stars along the magnitude 
axis. Finally, we fit a 2nd order polynomial to the resulting curves. As expected from 
the north-south reddening gradient, the fiducial lines in the north fields are significantly 
more reddened than those of the features, since the north fields are located more internally 
in the high dust-emission zone. Conversely, the south fields present fiducial lines very 
similar to those of the features, since these fields are also related to the cavity. The average 
reddening parameters of the \wA\ and \wB\ features, as well as those of the
north and south fields are summarized in Table~\ref{tab1}. Reddening parameters for the
symmetric field are also indicated (Sect.~\ref{nirp}).

\begin{table}
\caption[]{Average reddening properties}
\label{tab1}
\renewcommand{\tabcolsep}{0.90mm}
\begin{tabular}{cccccc}
\hline\hline
&\multicolumn{2}{c}{Blue sequence}&&\multicolumn{2}{c}{Red sequence}\\
\cline{2-3}\cline{5-6}\\
Feature&$\ejh$&${\rm A_V}$&&$\ejh$&${\rm A_V}$\\
\hline
North Field&$0.31\pm0.03$&$3.0\pm0.3$&&$1.07\pm0.07$&$10.3\pm0.7$\\
\wA&$0.18\pm0.02$&$1.7\pm0.2$&&$0.71\pm0.05$&$6.8\pm0.5$\\
South Field&$0.18\pm0.02$&$1.7\pm0.2$&&$0.71\pm0.05$&$6.8\pm0.5$\\
\hline
North Field&$0.29\pm0.03$&$2.8\pm0.3$&&$0.96\pm0.07$&$9.2\pm0.7$\\
\wB&$0.14\pm0.02$&$1.3\pm0.2$&&$0.68\pm0.05$&$6.5\pm0.5$\\
South Field&$0.14\pm0.02$&$1.3\pm0.2$&&$0.68\pm0.05$&$6.5\pm0.5$\\
\hline
Sym. Field&$0.24\pm0.03$&$2.3\pm0.3$&&$0.86\pm0.06$&$8.3\pm0.6$\\
\hline
\end{tabular}
\begin{list}{Table Notes.}
\item The values in this Table are based on the fiducial lines shown in 
Fig.~\ref{fig9}. We use the relation $\rm A_V=9.6\,\ejh$.
\end{list}
\end{table}

With the relation $\ebv=3\times\ejh$ (Sect.~\ref{mf}), the infrared reddening for \wA\ and \wB,
convert respectively to $\ebv=0.54$, and 0.42. These values are comparable to that of the Baade's 
window, $\ebv=0.47$ (Terndrup \cite{Tern1988}). Near the edge of Baade's window, the reddening 
increases to $\ebv=0.55$, in the direction of the globular cluster NGC\,6528 (Ortolani, Bica, 
\& Barbuy \cite{OBB1992}).

\section{The open cluster candidate Ru\,166}
\label{ru166}

Ru\,166 is a small stellar concentration located within the boundaries of the low-absorption 
window {$\rm W\,306.1-0.9$} (Sect.~\ref{nirphot}). Thus, it is important to investigate whether 
it is a small-scale window or dust hole, or a star cluster. In view of this we
will explore this object by means of tools usually applied to actual star clusters for the
 near-infrared range (e.g. Bonatto \& Bica \cite{BB2003}), and see whether Ru\,166 satisfies 
conditions to be an open cluster.

Ru\,166 is listed in catalogues of open clusters and open cluster candidates, e.g. Alter,
Ruprecht, \& Vanisek (\cite{Alter1970}), Lauberts (\cite{Lau1982}), Lyng\aa\ (\cite{Lynga1987})
and Dias et al. (\cite{Dias2002}). However, based on optical CCD photometry of individual stars
in a region $4\arcmin\times4\arcmin$, 
Piatti \& Clari\'a (\cite{PC2001}) indicated that Ru\,166 is not an open cluster. Instead, they 
suggest that the CMD morphology of Ru\,166 may arise from a combination of disk stars affected
by varying amounts of interstellar reddening.

\begin{figure*}
\begin{minipage}[b]{0.50\linewidth}
\includegraphics[width=\textwidth]{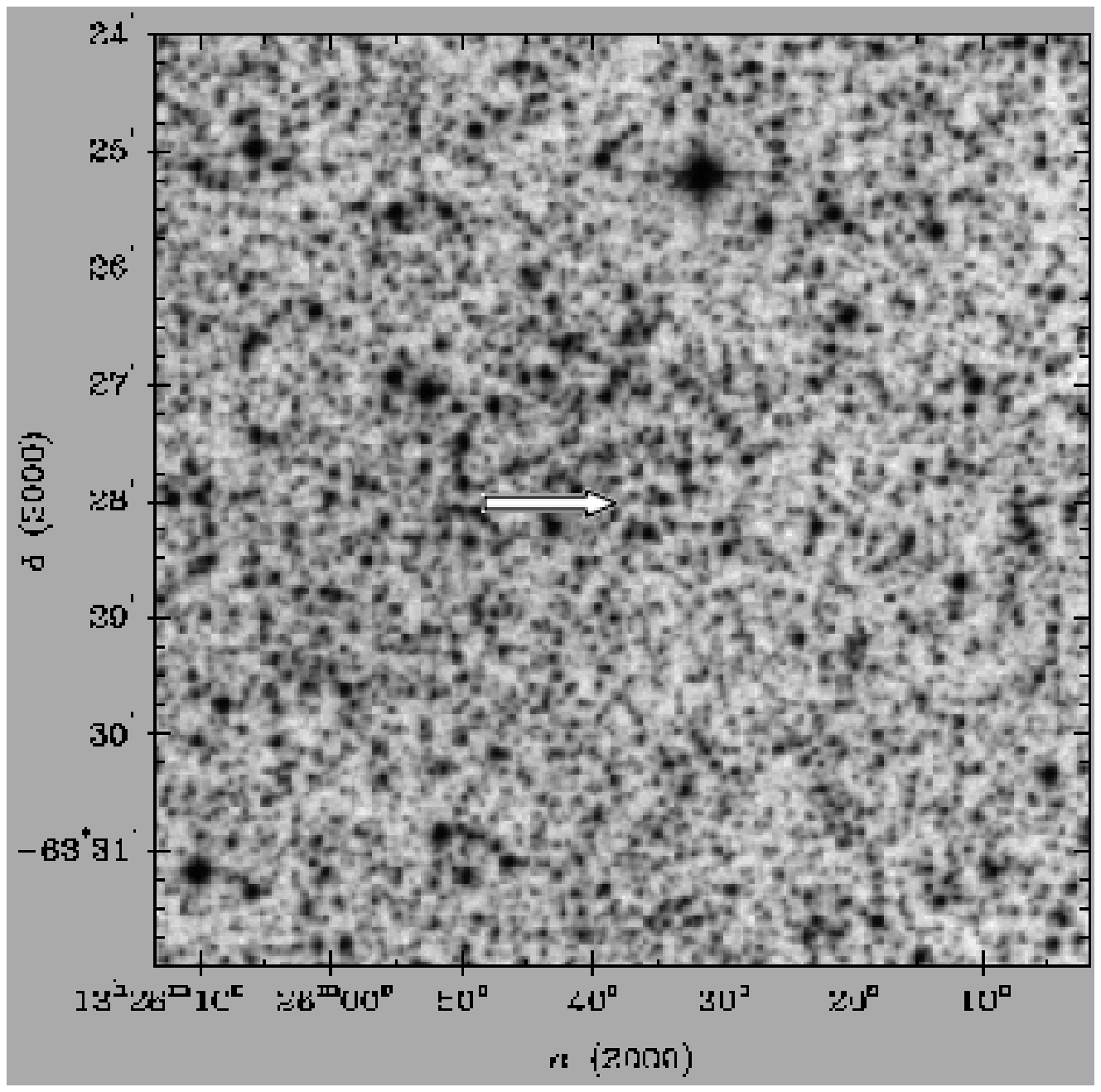}
\end{minipage}\hfill
\begin{minipage}[b]{0.50\linewidth}
\includegraphics[width=\textwidth]{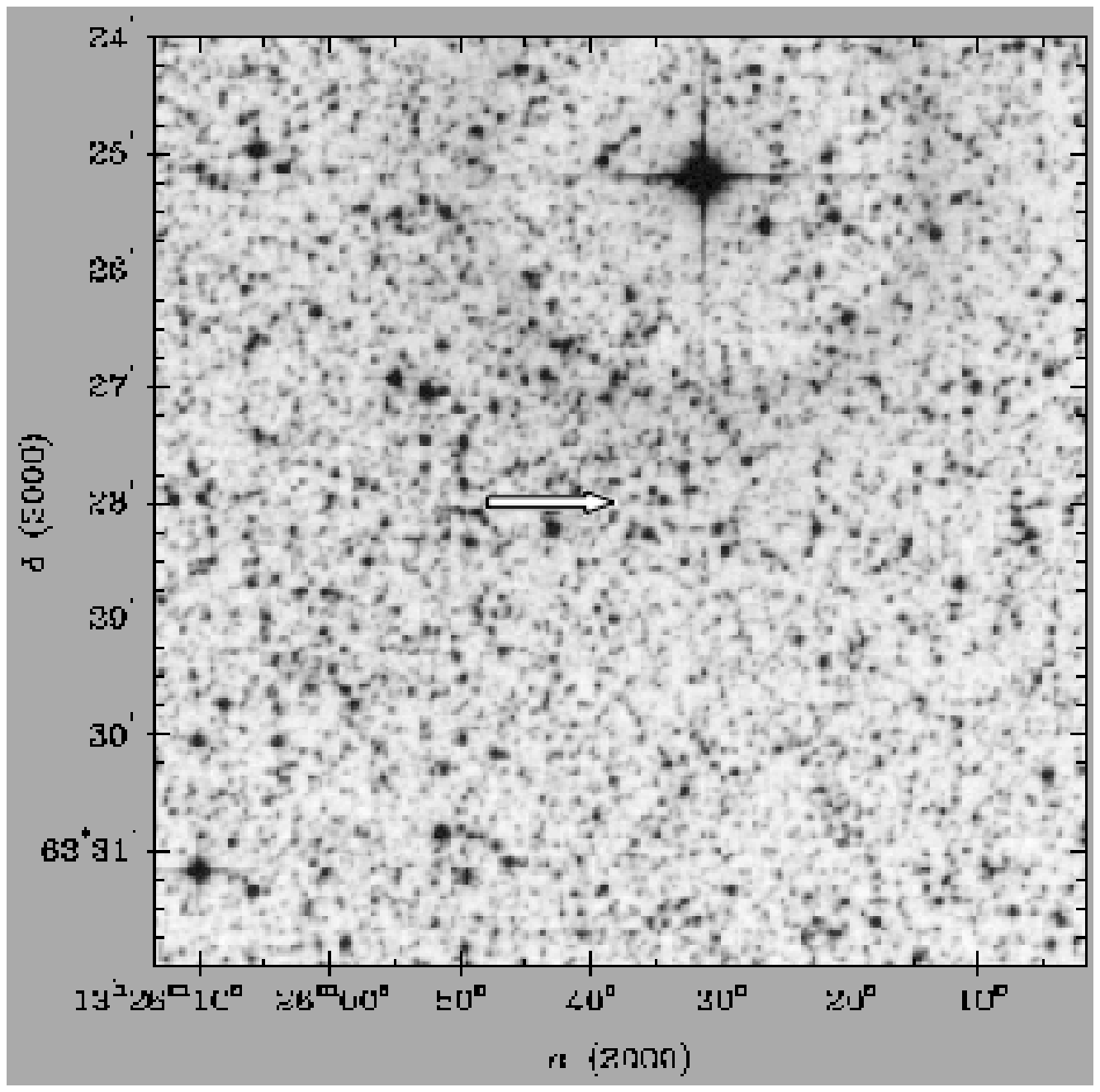}
\end{minipage}\hfill
\caption[]{Left panel: XDSS R image. Right panel: XDSS I image. The area covered in
both images is $8\arcmin\times8\arcmin$. The optical concentration is slightly off-centered
since the adopted center was derived from near-infrared star counts.  Arrows indicate
the location of the IR density peak.}
\label{fig10}
\end{figure*}

\begin{figure*}
\begin{minipage}[b]{0.54\linewidth}
\includegraphics[scale=1.1,width=\textwidth]{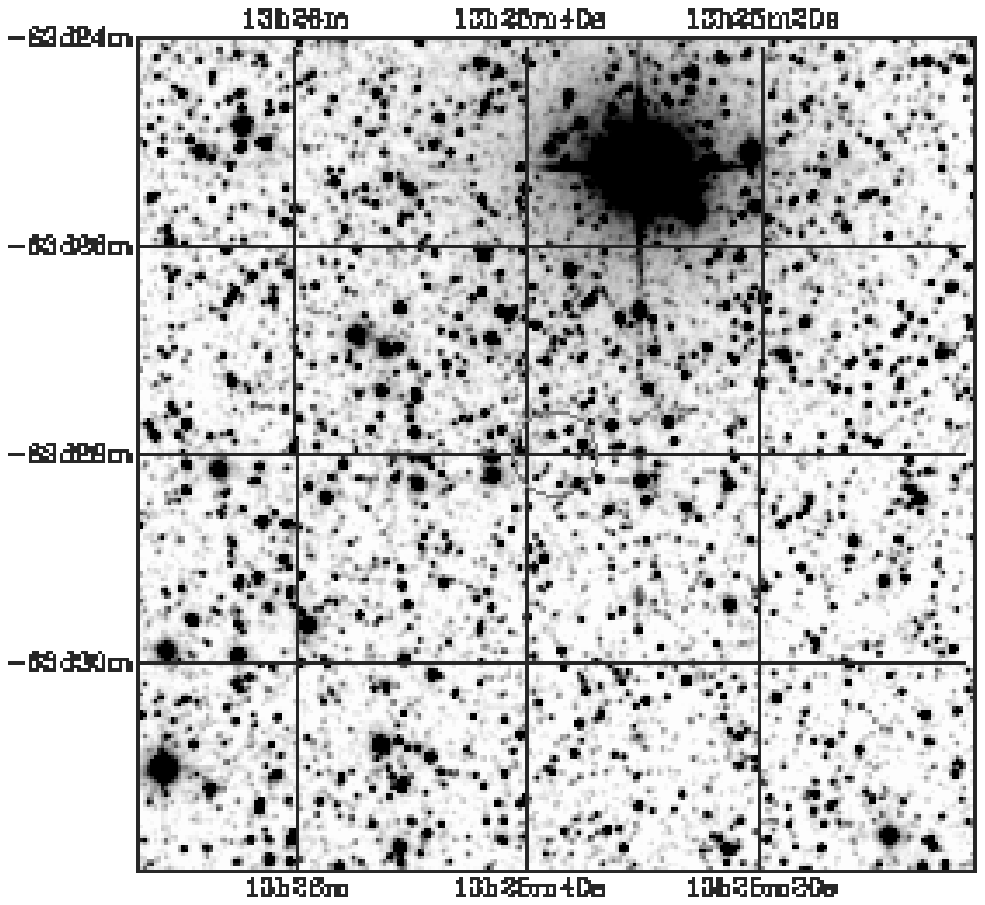}
\end{minipage}\hfill
\begin{minipage}[b]{0.46\linewidth}
\includegraphics[scale=1.1,width=\textwidth]{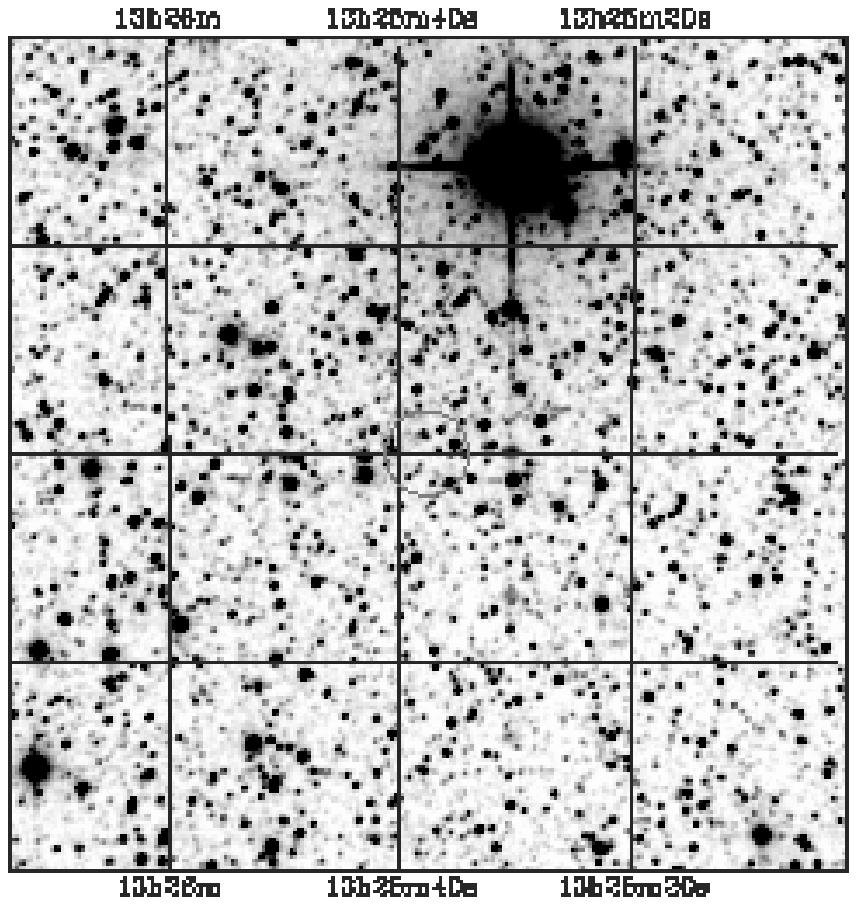}
\end{minipage}\hfill
\caption[]{Left panel: 2MASS H image. Right panel: 2MASS \ks\ image. The area covered 
in both images is $8\arcmin\times8\arcmin$. The near-infrared center is used as in 
Fig~\ref{fig10}.}
\label{fig11}
\end{figure*}

\begin{figure} 
\resizebox{\hsize}{!}{\includegraphics{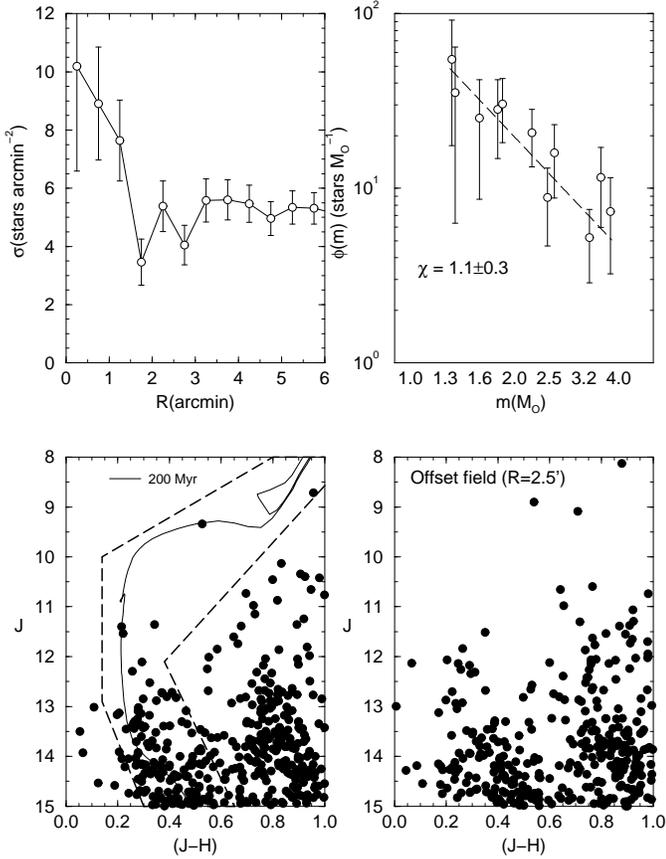}}
\caption[]{The open cluster candidate Ru\,166. Bottom left panel: CMD for the stars in the
central ($r\leq 2.5\arcmin$) with the 200\,Myr isochrone superimposed; parameters of the fit
are $\ejh=0.24\pm0.03$ and $\mMJ=12.0\pm0.1$, which correspond to $\ebv=0.76\pm0.1$
and $\ds=1.9\pm0.3\,$kpc. The dashed line marks off the colour-magnitude filter used to 
derive the radial density of stars and mass function. Top left: Radial distribution of stars; 
within uncertainties, the stellar concentration reaches up to $\approx2\,\arcmin$. Bottom 
right: CMD of the corresponding (same area) offset field. Top right: derived mass function.}
\label{fig12}
\end{figure}

Considering the stellar distribution in the optical images (Fig.~\ref{fig10}), the 
central position of Ru\,166 would be located at J2000.0 $\alpha=13^h25^m41^s$ and 
$\delta=-63^\circ\,27\arcmin\,25\arcsec$. However, the radial distribution of stars
(built for the stars with 2MASS photometry) centered on these coordinates does not produce 
a profile with the maximum value occurring for $r=0\arcmin$. Having this in mind, we have 
searched for new coordinates which maximize the central distribution of stars. We do this
by scanning in spatial detail the stellar density distribution in right ascension and 
declination. This clearly corresponds to a near-infrared peak in the stellar density 
distribution. The new central coordinates are J2000.0 $\alpha=13^h25^m38^s$ and 
$\delta=-63^\circ\,27\arcmin\,59\arcsec$, which correspond to $\ell=306.74^\circ$ and 
$b=-0.85^\circ$. Indeed, the shifts of 3\arcsec\ to the west in right ascension and 
34\arcsec\ to the south in declination, do produce a radial profile with the maximum 
at $r=0\arcmin$, as can be seen in Fig.~\ref{fig12} (top-left panel). Differences in dust 
attenuation between the optical and near-infrared may account for this small 
shift in the central position of Ru\,166.

Ru\,166, although generally located in a low-absorption zone, is found in a slight 
dip with respect to the distribution of the number-density of stars (Fig.~\ref{fig4}). 
However, the sampling used to build that number-density distribution of stars corresponds 
to 12\arcmin, while the spatial scale of Ru\,166 is about 2\arcmin. Thus, the distribution 
shown in Fig.~\ref{fig4} is not sensitive to small-scale fluctuations in the number-density 
of stars. Consequently, we must explore Ru\,166 in more spatial detail than in the map of Fig.~\ref{fig4}. 
In Fig.~\ref{fig10} we provide $R$ and $I$ XDSS images of an $8\arcmin\times8\arcmin$ 
region centered on Ru\,166, extracted from the Canadian Astronomy Data Centre (CADC\footnote{\em 
http://cadcwww.dao.nrc.ca/}). As expected, there is a slight excess of stars in the center of 
both images. However, when the same region is considered in the near-infrared, the stellar 
surface density becomes more uniform, as can be seen in the 2MASS H (left panel of Fig.~\ref{fig11}) 
and \ks\ (right panel) images. One explanation for this effect may be that, since the near-infrared 
is less affected by dust attenuation than the optical, the stellar surface density of the background 
increases uniformly throughout a given region. Thus, a small cluster with a low stellar surface 
density, particularly in the outer regions, may get  overwhelmed in the background. This effect
 would be particularly critical for small clusters projected against populous backgrounds 
such as the disk or bulge.

To investigate the nature of this stellar concentration, we extracted 2MASS data in a region 
with 10\,\arcmin\ in radius centered on the above coordinates. Note that our analysis employs 
a larger area than that sampled by Piatti \& Clari\'a (\cite{PC2001}). The CMD for 
the stars in the central 2.5\,\arcmin\ is shown in the bottom-left panel of Fig.~\ref{fig12}, 
while in the bottom-right panel we show the CMD of the comparison field (same area), which is 
taken from an external annulus. At first sight both CMDs are similar. Although the stars are 
sparsely distributed 
throughout the CMD, we still can tentatively apply fits with isochrones to estimate parameters 
for Ru\,166. The {\em best-fit} has been obtained with the 200\,Myr, solar-metallicity Padova 
isochrone, $\ejh=0.24\pm0.03$ and $\mMJ=12.0\pm0.1$, which correspond to $\ebv=0.76\pm0.1$ and 
a distance to the Sun $\ds=1.9\pm0.3\,$kpc, thus slightly beyond the Sgr-Car arm in this 
direction.

We also built a CMD with the same area as above with extracted around the optical center
discussed above for Ru\,166. The resulting CMD was essentially the same as that built
around the near-infrared center (Fig.~\ref{fig12}, bottom-left panel).

With respect to the radial distribution of stars (top-left panel in Fig.~\ref{fig12}),
Ru\,166 presents a central concentration of stars, reaching up to $\approx2\,$\arcmin.
The dips in the radial profile at $\approx2\arcmin$ and $\approx3\arcmin$ are suggestive of dust.
With a distance to the Sun of $\ds\sim1.9$\,kpc, the linear diameter of Ru\,166 would
be $\approx2\,$pc, which would put this object in the small-cluster tail, with respect 
to the open cluster size distribution (Binney \& Merrifield \cite{Binney1998}).  Recall
that, in order to maximize membership in the case of a star cluster, the above analysis was 
restricted to the stars inside a colour-magnitude filter (dashed line in the bottom-panel of 
Fig.~\ref{fig12}). For a discussion on the use of colour-magnitude filters
in the analysis of star clusters, see e.g. Bonatto, Bica, \& Santos Jr. (\cite{BBJ2004}).
The central concentration of stars presents a relatively high density-contrast 
with respect to the stellar density in the surrounding regions. However, due to the small-scale 
structure of the radial distribution of stars, a King model could not be fitted to the profile 
of Ru\,166. Besides dust, the small-scale structure of Ru\,166 may be accounted for the high
density of stars in the area where it is located, ${\rm\sim28\,stars\,arcmin^{-2}}$ (Fig.~\ref{fig4}),
which probably would drown any extended structure of a cluster. Indeed, the density of stars
for $r\geq3\arcmin$ around the center of Ru\,166 still remains rather high, ${\rm\sim5\,stars\,arcmin^{-2}}$, 
after applying the colour-magnitude filter (top-left panel in Fig.~\ref{fig12}). We remark that
the filtering process increases the contrast so that Ru\,166 has a stellar density of a factor
of $\sim2$ as compared to that of the background (top-left panel in Fig.~\ref{fig12}). This
might suggest the presence of an open cluster. As a comparison, the contrast between cluster and
background for the CMD unfiltered star sample for M\,67 reaches a factor of $\sim5$ (Bonatto
\& Bica \cite{BB2003}).

Finally, in the case of a cluster, we can adopt the 200\,Myr isochrone fit as representative 
of the stars in Ru\,166 to calculate the overall mass function ($\phi(m)=\frac{dN}{dm}$) in 
this stellar concentration. To represent the offset field we use stars in the annulus between 
5\arcmin\ and 10\arcmin. The colour-magnitude filter (bottom-panel of Fig.~\ref{fig12}) has 
been applied to the stars both in Ru\,166 and offset field. The result is shown in the top-right 
panel of Fig.~\ref{fig12}. A fit with the function $\phi(m)\propto m^{-(1+\chi)}$ resulted 
in a slope $\chi=1.1\pm0.4$ which, within uncertainties, is similar to a standard Salpeter 
($\chi=1.35$) mass function. The observed MS mass (in the range $1.3\leq m(\ms)\leq3.8$) 
results in $\mobs=81\pm34\,\ms$, with a number of observed MS stars of $\approx39$. An 
estimate of the total mass locked up in stars in Ru\,166 can be made by taking into
account all stars down to the low-mass limit, $0.08\,\ms$. We do this by assuming the 
universal IMF of Kroupa (\cite{Kroupa2001}), in which $\chi=0.3\pm0.5$ for $0.08\,\ms 
- 0.5\,\ms$. For masses in the range $0.5\,\ms - 1.3\,\ms$ we use the presently derived
value of $\chi$. The total (extrapolated) mass results $\mtot=259\pm98\,\ms$, and the 
corresponding average mass density in Ru\,166 becomes ${\rm\rho=49\pm18\,\ms\,pc^{-3}}$.

The data and analysis presented in the paragraphs above do not allow us to draw a conclusive 
diagnosis on the nature of Ru\,166. On the one hand, the presence of a stellar 
concentration as shown in Figs.~\ref{fig10} and \ref{fig12} (top-left panel), and the 
$\chi\approx1.1$ mass function Fig.~\ref{fig12} (top-right panel) are suggestive of an open 
cluster. On the other hand, the similarity between the CMDs of Ru\,166 and offset field (bottom 
panels in Fig.~\ref{fig12}) as well as the low contrast in the stellar surface density in the 
near-infrared (Fig.~\ref{fig11}) suggest that Ru\,166 may result from a fluctuation in the 
number-density of field stars probably associated  with a small scale low-absorption region. 
We conclude that further data, particularly proper motion, are needed to  definitively pin 
down the nature of Ru\,166.

\section{Concluding remarks}
\label{conclu}

This paper dealt with the nature of two conspicuous stellar concentrations in
Centaurus, \wA\ at $\ell=306.47^{\circ}$, $b=-0.61 ^{\circ}$, and 
\wB\ at $\ell=307.01^{\circ}$, $b=-0.74 ^{\circ}$, projected against the
disk. By means of 2MASS infrared photometry, we have found a reddening gradient
perpendicular to the galactic plane, characterizing an overall low-latitude, 
low-absorption zone towards the disk. Although we are certainly dealing with stellar 
populations with spatial depth and age mixtures, Padova isochrone fits to the bulk of the CMDs 
indicate that stars in both structures are dominantly located at $\sim1.5$\,kpc from 
the Sun, a distance consistent with that of the Sgr-Car arm. The presence of this low-absorption 
window provides a means to probe the Sgr-Car arm itself, and more distant parts of the disk, 
probably the tangent part of the Sct-Cru arm, and/or the far side of the 
Sgr-Car arm. The existence of low-absorption zones is important for
deep optical studies of the Galaxy. We also investigated the nature 
of the open cluster candidate Ru\,166. In some aspects (e.g. the stellar concentration 
in the optical and the $\chi\approx1.1$ mass function) Ru\,166 resembles an 
open cluster. However, the similarity between the CMDs of Ru\,166 and offset field,
as well as the null-contrast in near-infrared images suggest that Ru\,166 may result
from a fluctuation in the number-density of field stars probably associated  with
a small scale low-absorption region. Thus, more data, proper motion in particular, are 
needed to decide upon this issue.

Throughout this work we applied several methods to test whether we were dealing
with windows and dust holes or stars clusters and associations. They included
CMD morphology, radial distribution of stars and number count of stars as a function
of magnitude, both for the optical and near-infrared. The features \wA\ and \wB\ do not 
appear to qualify as star clusters, since they do not pass all of the above tests.

Finally, the present study showed significant variations in the stellar surface density 
in different spatial scales in optical images. However, in the near-infrared the stellar 
surface density contrast ($\Delta\sigma/\sigma$) is nearly absent. One possibility 
is non-uniform dust attenuation, which could be produced by a small dust hole or a window. 
Another possibility is intrinsic stellar distribution variations, which would become more 
uniform in a near-infrared image owing to the increased background density associated to 
less dust attenuation. Stellar concentrations, in this case, would be actual star clusters, 
associations or star-forming regions. A poorly populated star cluster will stand out against 
a sparse field. However, when projected against the disk, it will present a lower contrast
$\Delta\sigma/\sigma$. In order to disentangle the effects above it would require 
detailed modelling of Galactic structures in the near-infrared, including spiral arms. We 
intend to work in this direction in a future paper.

\begin{acknowledgements}
 We thank the referee Dr. Cameron Reed for interesting comments which improved the paper.
This publication makes use of data products from the Two Micron All Sky Survey, which is a 
joint project of the University of Massachusetts and the Infrared Processing and Analysis 
Center/California Institute of Technology, funded by the National Aeronautics and Space 
Administration and the National Science Foundation. We employed catalogues from CDS/Simbad 
(Strasbourg) and  Digitized Sky Survey images from the Space Telescope Science Institute 
(U.S. Government grant NAG W-2166) obtained using the extraction tool from CADC (Canada). 
We acknowledge support from the Brazilian Institution CNPq. The Guide Star Catalogue-II 
is a joint project of the Space Telescope Science Institute and the Osservatorio Astronomico 
di Torino. Space Telescope Science Institute is operated by the Association of Universities 
for Research in Astronomy for the National Aeronautics and Space Administration under 
contract NAS5-26555. The participation of the Osservatorio Astronomico di Torino is supported 
by the Italian Council for Research in Astronomy. Additional support is provided by
European Southern Observatory, Space Telescope European Coordinating Facility, the 
International GEMINI project and the European Space Agency Astrophysics Division.
\end{acknowledgements}

%
%


\begin{thebibliography}{}

\bibitem[1970]{Alter1970}
   Alter, G., Ruprecht, J., \& Vanisek, J. 1970, in {\em Catalogue of Star
   Clusters and Associations}, ed. G. Alter, B. Bal\'azs, \& J. Ruprecht
   (Akademiai Kiado, Budapest)

\bibitem[1946]{Baade1946} 
   Baade, W. 1946, PASP, 58, 249
   
\bibitem[1963]{Baade1963} 
   Baade, W. 1963, in {\em Evolution of stars and galaxies}, Harvard Univ. Press,
   Cambridge, Mass., 277   
   
\bibitem[2003]{BiBoDu2003} 
   Bica, E., Bonatto, C., \&  Dutra, C.M. 2003, A\&A, 405, 901
   
\bibitem[2004]{BiBoDu2004} 
   Bica, E., Bonatto, C., \&  Dutra, C.M. 2004, A\&A, in press 
   
\bibitem[1998]{Binney1998} 
   Binney, J., \& Merrifield, M. 1998, in {\em Galactic Astronomy}, Princeton, 
   NJ: Princeton University Press. (Princeton series in astrophysics) 
   
\bibitem[2003]{BB2003} 
   Bonatto, C., \& Bica, E. 2003, A\&A, 405, 525
   
\bibitem[2004]{BBJ2004} 
   Bonatto, C., Bica, E., \& Santos Jr, J.F.C. 2004, A\&A, submitted
   
\bibitem[2004]{BBG2004} 
   Bonatto, C., Bica, E., \& Girardi, L. 2004, A\&A, 415, 571

\bibitem[2002]{Dias2002}
   Dias, W.S., Alessi, B.S., Moitinho, A. \& L\'epine, J.R.D. 2002, A\&A, 389, 871

\bibitem[2003]{DCS2003} 
   Daflon, S., Cunha, K., Smith, V.V. \& Butler, K. 2003, A\&A, 399, 525

\bibitem[2001]{DCS2001} 
   Daflon, S., Cunha, K., Butler, K. \& Smith, V.V. 2001, ApJ, 563, 325

\bibitem[1888]{Dreyer1888} 
   Dreyer, J.L.E. 1888, MmRAS, 49, 1

\bibitem[2002]{DSB2002} 
   Dutra, C.M., Santiago, B.X. \& Bica, E. 2002, A\&A, 381 219

\bibitem[2002]{DB2002}
   Dutra, C.M., \& Bica, E. 2002, A\&A, 383, 631
   
\bibitem[1968]{Fitz68}
   Fitzgerald, M.P. 1968, AJ, 73, 983

\bibitem[1976]{GG1970}
   Georgelin, Y.P., \& Georgelin, Y.M. 1970, A\&AS, 3, 1

\bibitem[1976]{GG1976}
   Georgelin, Y.M., \& Georgelin, Y.P. 1976, A\&A, 49, 57
   
\bibitem[2002]{Girardi2002} 
   Girardi, L., Bertelli, G., Bressan, A., et al. 2002, A\&A, 391, 195
   
\bibitem[1962]{King1962}
   King, I. 1962, AJ, 67, 471   
   
\bibitem[2001]{Kroupa2001}
   Kroupa, P. 2001, MNRAS, 322, 231 
   
\bibitem[1982]{Lau1982}   
   Lauberts, A. 1982, in {\em ESO/Uppsala survey of the ESO(B) atlas}, Garching: 
   European Southern Observatory (ESO)

\bibitem[1976]{LLE1976} 
   Lloyd Evans, T. 1976, MNRAS, 174, 169

\bibitem[1987]{Lynga1987}
   Lyng\aa, G. 1987, in {\em Catalogue of Open Cluster Data}, Strasbourg:
   Centre de Donn\'ees Stellaires
   
\bibitem[2002]{Mathys2002} Mathys, G., Andrievsky, S.M., Barbuy, B., Cunha, K. \& 
   Korotin, S.A. 2002, A\&A, 387, 890

\bibitem[1994]{MWR1994} 
   McWilliam, A. \& Rich, R.M. 1994, ApJS, 91, 749
   
\bibitem[1992]{OBB1992}
   Ortolani, S., Bica, E., \& Barbuy, B. 1992, A\&AS, 92, 441
   
\bibitem[2001]{PC2001} 
   Piatti, A.E., \& Clari\'a, J.J. 2001, A\&A, 379, 453

\bibitem[2003]{Rus2003}
   Russeil, D. 2003, A\&A, 397, 133
   
\bibitem[1976]{Schw76}
   Schweizer, F. 1976, ApJS, 31, 313
   
\bibitem[1997]{2mass} Skrutskie, M., Schneider, S.E., Stiening, R., et al. 1997, in {\em 
    The Impact of Large Scale Near-IR Sky Surveys}, ed. Garzon et al., Kluwer (Netherlands), 
    210, 187
  
\bibitem[1988]{Tern1988} 
   Terndrup, D.M. 1988, AJ, 96, 884
   
\bibitem[2001]{Vera2001}
   Vera-Villamizar, N., Dottori, H., Puerari, I. \& de Carvalho, R. 2001, ApJ, 547, 187   

\end{thebibliography}
\end{document}